\documentstyle[twocolumn,aps]{revtex}
\input epsf.sty
\newtheorem{proposition}{Proposition}
\begin{document}
\wideabs{
\title{Solution of the Density Classification Problem with Two Cellular
Automata Rules}
\draft
\author{Henryk Fuk\'s}
\address{Department of Physics, University of Illinois, Chicago,
IL 60607-7059\\ { fuks@uic.edu}}
\maketitle
\begin{abstract}
Recently, Land and Belew {[}Phys. Rev. Lett. {\bf 74}, 5148 (1995){]} have
shown that no one-dimensional two-state cellular automaton which
classifies binary
strings according to their densities of 1's and 0's can be constructed.
We show that a pair of elementary rules, namely the ``traffic rule'' 184
and the ``majority rule'' 232, performs the task perfectly. This solution
employs the second order phase transition between the freely moving
phase and the jammed phase occurring in rule 184. We present exact
calculations of the order parameter in this transition using the method
of preimage counting.
\end{abstract}
}

\section{Introduction}
%\thispagestyle{myheadings}
%\markboth{}{\rm To appear in Phys. Rev. E {\bf 55}, March 1997}
In recent years, cellular automata (CA)
\cite{Wolfram94} have received
considerable attention as
models of natural systems in which simple local interactions between
components give rise to a complex global behavior. Such systems
have the ability of coordinated global
 information processing, often called ``emergent computation,'' which could not
be achieved by a single component.
Since emergent computation occurs in many biological systems such as
the brain, the immune system, or insect colonies, it is natural to ask how
the evolution produces such complex information processing
capabilities in
ensembles of simple locally interacting elements.

To model this process, genetic algorithms have been used to evolve
cellular automata capable of performing specific computational tasks, in
particular the so-called density classification task \cite{MHC93}.
 The CA performing this task
should converge to a fixed point of all 1's if the initial configuration
contains more 1's than 0's, and to a fixed point of all 0's if the initial
configuration contains more 0's  than 1's. This should happen within $M$ time
steps, where, in general, $M$ can depend on the lattice size $L$
(assuming periodic boundary conditions).

The earliest proposed solution to this problem was the two-state radius-3
rule constructed by Gacs, Kurdyumov, and Levin (GKL) \cite{GKL78}. According to
this rule, if the state of a cell is 0, its new state is determined
by a majority vote among itself, its left
neighbor, and its second left neighbor. If the state of the cell is 1,
its new state is given by the majority vote among itself, its right
neighbor, and its second right neighbor. It has been demonstrated that
the GKL rule performs the density classification task only approximately, i.e.,
not all initial configurations are classified correctly. In particular,
when the initial
density is close to $1/2$, approximately $30\%$ of the initial configurations
are misclassified.
Attempts to evolve CA that perform density classification task
resulted in rules comparable to GKL in terms of proficiency, but not better,
typically classifying correctly about $80\%$ of all possible
initial configurations \cite{MHC93}.
In fact, it has been recently proved by Land and Belew \cite{LB95} that
the perfect two-state rule performing this task does not
exist.

If we think about the cellular automaton as a model of a multicellular
organism composed of identical cells, or a single kind of ``tissue,''
we could say that evolution reached a ``dead end'' here.
In the biological evolution, when the single-tissue organism cannot be
improved any further, the next step is the differentiation of cells,
or aggregation of cells into organs adapted to perform a specific function.
In a colony of insects, this can be interpreted as a ``division of labor,''
when separate groups of insects perform different partial tasks.
For cellular automata, this could be realized as an ``assembly line,''
with two (or more) different CA rules: the first rule is iterated $t_1$
times, and then the resulting configuration is processed by another rule
iterated $t_2$ times. Each rule plays the role of a separate ``organ,''
thus we can expect that such a system will be able to perform complex
computational tasks much better that just a single rule.
In what follows, we will show that for
the density classification problem this is indeed the case,
and that the perfect performance can be achieved with just two elementary
rules (184 and 232) arranged in the ``assembly line'' described above.

\section{Rule 184}

Let ${\cal G}=\{0,1\}$ be called {\it a symbol set}, and let $\cal S$
 be the set of all bisequences over ${\cal G}$, where by a bisequence we mean a
 function on  $\mathbf{Z}$ to ${\cal G}$. The set  ${\cal S}$ will be
called {\it the configuration space}.
{\it A block of length} $n$ is an ordered set $b_{1} b_{2} \ldots b_n$,
where $n\in {\mathbf{N}}$, $b_i \in {\cal G}$. ${\cal B}_n$ denotes the set
of all blocks of length $n$. The number of elements of ${\cal B}_n$
(denoted by ${\mathop{\rm{card\,}}} {\cal B}_n$) equals $2^{n}$.

A mapping $f:\{0,1\}^{3}\mapsto\{0,1\}$ will be called {\it an elementary
 cellular automaton rule}. Alternatively, the function $f$ can be
 considered as a mapping of ${\cal B}_3$ into ${\cal B}_0={\cal G}=\{0,1\}$.
Corresponding to $f$ (also called {\it a local mapping}) we define a
 {\it global mapping}  $F:S\to S$ such that
$
[F(s)]_i=f(s_{i-1},s_i,s_{i+1})
$
 for any $s\in S$.

A {\it block evolution operator} corresponding to the local rule $f$
is a mapping
 ${\rm \bf f}:{\cal B}_n \mapsto {\cal B}_{n-2}$  defined as
\begin{eqnarray}
{\rm \bf f}(b_1 b_2 &\ldots& b_n)= \\ \nonumber
& &f(b_{1},b_{2},b_{3}) f(b_{2},b_{3},b_{4})
\ldots f(b_{n-2},b_{n-1},b_{n}),
\end{eqnarray}
 where $n > 2$.

Rule 184 has been studied as a simplest model for the road traffic
flow \cite{fukui95} .
Its rule table
\begin{eqnarray*}
& & 000 \rightarrow 0,\, 001 \rightarrow 0,\, 010 \rightarrow 0,\,
011 \rightarrow 1, \\ & & 100 \rightarrow 1,\, 101 \rightarrow 1,\,
110 \rightarrow 0,\, 111 \rightarrow 1
\end{eqnarray*}
can be interpreted in terms of ``cars'' (ones)  and ``empty spaces''
(zeros). ``Cars'' are moving to the right. If the ``car'' has an
``empty space'' in front of it, it will move there, i.e., it will move
one unit to the right. Under this rule, the number of ``cars'' does not
change, or in other words, the density of 1's is conserved.

In order to understand the dynamics of rule 184,
let us define a {\em preimage} of a finite block $a\in {\cal B}_k$ as a block
$b \in {\cal B}_{k+2}$ such that ${\rm \bf f}(b)=a$. Similarly, a {\em n-step preimage}
of the block $a \in {\cal B}_{k}$ is a block $b \in {\cal B}_{k+2n}$ such that
${\rm \bf f}^{n}(b)=a$, where the superscript $n$ denotes multiple composition
of ${\rm \bf f}$, i.e., iterating ${\rm \bf f}$ $n$-times. For a given elementary rule, the
number of $n$-step preimages of a given block $b$ (we will denote this number
by ${\mathop{\rm{card\,}}} {\rm \bf f}^{-n}(b)$)
 can be anything from $0$ to
$2^{2+2n}$. For many rules, an exact expression for
${\mathop{\rm{card\,}}} {\rm \bf f}^{-n}(b)$ can
be found \cite{fuks3}, and the ``traffic'' rule 184
is among such exactly solvable cases.

For convenience, let us consider the preimages of the block
$00$ under ${\rm \bf f}_{184}$. We will first prove the following proposition:
\begin{proposition}
The block $s_1 s_2 \ldots s_{2n+2}$ is an $n$-step preimage of $00$ under rule
$184$ if and only if
$s_1=0$, $s_2=0$ and $2+\sum_{i=3}^{k} \xi(s_i) > 0$ for every
$3 \leq k \leq 2n+2$, where $\xi(0)=1$, $\xi(1)=-1$.
\end{proposition}
%%%%%%%%%%%%%%%%%%%%%%%%%%%%%%%%%%%%%%%%
We will present only a sketch of the proof here, based on the concept
of ``defects'' \cite{bnr90,bnr91,Hanson92,Crutchfield93}. Since the dynamics of rule 184
can be viewed as ``particles'' or ``defects'' (blocks of two or more 0's
or 1's) propagating
in the regular background (periodic pattern of alternating 1 and 0,
$\ldots 101010101 \ldots$), it will be useful to introduce a transformation
eliminating the background.
Let us consider a block of length $2n+2$,
and let us check whether it is a preimage of $00$ or not. To eliminate
the background,
we first identify all continuous clusters of at least two zeros. Each site in
 such a cluster
is replaced by the symbol $\alpha$, except the leftmost 0, which
is replaced by $\star$. Similarly, in every cluster of at least two ones,
every 1 is replaced by the symbol $\beta$, except the rightmost 1, which
is replaced by $\star$. All remaining sites are replaced by $\star$.
 For example, the string
\mbox{$1010001011000010$} will be transformed into
\mbox{$\star\star\star\star\alpha\alpha\star\star\beta\star\star\alpha\alpha
\alpha\star\star$}. The dynamics of the rule can now be understood as a
motion of blocks of $\alpha$'s and $\beta$'s in the background of
$\star$'s. Every time step, each
block of $\alpha$'s moves one unit to the right, and each
block of $\beta$'s one unit to the left. When $\alpha$ and $\beta$ blocks
collide,  each block
decreases its length by one per time step, until one of them (or both)
disappear. Let us now consider the block $s_1s_2 \ldots s_{2n+2}$
iterated $n$ times with rule 184. The state of the cell $i$ at
time $t$ will be denoted by $s_i^t$. Block $s_1^0 s_2^0 \ldots s_{2n+2}^0$
can be a preimage of $00$ when $s_n^ns_{n+1}^n=00$, or using our transformation,
$s_n^ns_{n+1}^n=\star\alpha$. Since all blocks of $\alpha$'s are moving with
constant speed, this means that at $t=0$ we must have
$s_1^0 s_2^0=\star\alpha$ or, using the original notation,
$s_1^0 s_2^0=00$, which means that $\alpha$ travels from $i=2$
to $i=n+1$ in $n$ time steps. It can travel this distance ``safely,''
if and only if it does not collide with another $\beta$-block. This can
happen if all $\beta$-blocks are annihilated before they hit our $\alpha$,
or in other words, if for every $k$ such that $2\leq k \leq 2n+2$ the number of $\beta$'s
in the subblock $s_3^0s_4^0 \ldots s_{k}^0$ is smaller than
the number of $\alpha$'s. Translating this back to the original notation
(i.e., using 0's and 1's) we obtain the conclusion of Proposition 1.
%%%%%%%%%%%%%%%%%%%%%%%%%%%%%%%%%%%%%%%%
% End of proof
%%%%%%%%%%%%%%%%%%%%%%%%%%%%%%%%%%%%%%%%

The $s_3s_4 \ldots s_{2n+2}$ part of the preimage of $00$ can therefore be
constructed by using the following algorithm: we start with an
initial ``capital'' equal to 2. Every time we choose 0, our ``capital''
increases by 1, and when we choose 1, it decreases by 1. We have to find a
path such that the ``capital'' never reaches zero -- the problem
known in the probability theory as the ``gambler's ruin problem'' \cite{feller}.
Let us denote the ``capital'' at time $t$ by
$c(t)=\sum_{i=3}^{t} \xi(s_i)$, and let $c(2)=2$. Our string
$s_3s_4 \ldots s_{2m+2}$ can be now represented by a path
$c(2)c(3) \ldots c(2m+2)$. Geometrically, this can be viewed as a
two-dimensional
polygonal line joining points $[t,c(t)]$ starting at $(2,2)$ and
ending at $(2m+2,y)$ which neither touches nor crosses the horizontal axis.
The number of all such paths can be computed using well known
combinatorial theorems (see, for example, \cite{feller}) and is equal
to $N_{2m,y-2}-N_{2m,y+2}$, where
\begin{equation}
N_{a,b}=\left( \begin{array}{c}
                a \\ \frac{1}{2}(a+b)
               \end{array} \right),
\end{equation}
and $N_{a,b}=0$ if $b>a$.

In the path from $(2,2)$ to $(2m+2,y)$ we have $(2m+y-2)/2$ zeros
and $(2m-y+2)/2$ ones.
If we randomly choose 1's with probability $p$ and 0's with probability
$1-p$, then the probability of selecting admissible path with a given
``finite capital'' $y$ is
\begin{equation}
(N_{2m,y-2}-N_{2m,y+2}) p^{(2m-y+2)/2} (1-p)^{(2m+y-2)/2}
\end{equation}
Taking into account the fact that $y$ must be even and that the first
two digits of the preimage must be $00$, we conclude that the probability
that a randomly selected string (if we select ones with probability $p$
and zeros with probability $1-p$)
of length $2m+2$ is an $m$-step preimage
of $00$ is
\begin{eqnarray}
\label{resulteq}
P_m(00)=
(1-p)^2\sum_{k=1}^{m+1}(N_{2m,2k-2}&-& \\ \nonumber
-N_{2m,2k+2}) p^{m-k+1} (1-p)^{m+k-1}.
\end{eqnarray}
Note that if we start from an infinite random initial configuration
with the density of ones $p$, then the probability of the occurrence of the block
$00$ after $m$ iterations of rule $184$ will also be given
by $P_m(00)$.

Although eq.  (\ref{resulteq}) is rather complicated, asymptotic
expansion for large $t$ is possible. Using the Stirling formula to approximate
binomial coefficients, after some algebra one obtains:
\begin{equation}
P_t(00) \simeq \left\{ \begin{array}{ll}
 1-2p + p \displaystyle{\frac{[4p(1-p)]^t}{\sqrt{\pi t}}} & \mbox{if $p<\frac{1}{2}$}, \\ [1em]
 (1-p)\displaystyle{\frac{[4p(1-p)]^t}{\sqrt{\pi t}}}     & \mbox{otherwise},
\end{array}
\right.
\end{equation}
and therefore
\begin{equation}
P_{\infty}(00) = \left\{ \begin{array}{ll}
 1-2p  & \mbox{if $p<\frac{1}{2}$}, \\
 0    & \mbox{otherwise}.
\end{array}
\right.
\end{equation}
As we can see, $P_{\infty}(00)$ plays here a role of the order parameter
in a second order kinetic phase transition, with the control parameter $p$.
The critical point is exactly at $p=1/2$, and at the critical point
$P_t(00)$ approaches its stationary value as $t^{-1/2}$. Away from the
critical point, the approach is exponential, and it slows down as $p$ comes
closer to $1/2$. For finite configurations (with periodic boundary condition) the performance
of rule 184 in eliminating $00$ blocks for $p>1/2$ is even better.
\begin{proposition}
If the finite initial configuration consists of $N_0$ zeros and $N_1$ ones, and
$N_0<N_1$, then after at most $\lfloor (N_0+N_1-2)/2 \rfloor$ time steps all $00$ blocks
disappear ($\lfloor x \rfloor$ denotes the largest integer less or equal to $x$).
\end{proposition}
To see this, let us first consider $N_0+N_1$ even, so that $N_1-N_0 \geq 2$.
Let us further assume that after $(N_0+N_1-2)/2$ time steps we still
have at least one $00$ block. This means that we can write our entire initial
configuration as $a_1a_2 \ldots a_{N_0+N_1}$ satisfying hypothesis of Proposition~1,
i.e.,
$a_1=0$, $a_2=0$ and $2+\sum_{i=3}^{k} \xi(a_i) > 0$ for every k such that
$3 \leq k \leq N_0+N_1$. Note, however, that if $a_1=0$, $a_2=0$ then
$2+\sum_{i=3}^{N_0+N_1} \xi(a_i) \leq 0$, and since it contradicts the previous
statement,
$a_1a_2 \ldots a_{N_0+N_1}$ cannot be a preimage of $00$.
The proof for odd $N_0+N_1$ is similar. Also, due to the self-duality of rule $184$, the same theorem holds for the block $11$ when $N_0>N_1$.
When $N_0=N_1$, both $00$ and $11$ blocks disappear after $(N_0+N_1-2)/2$
time steps, and the configuration becomes an alternating sequence of 0 and 1,
$\ldots 01010101 \ldots$.

To summarize, we found that for a finite lattice of length $L$ and
the  density $\rho$,
after $\lfloor (L-2)/2 \rfloor$ iterations of rule 184 the resulting configuration
\begin{itemize}
\item contains no $00$ blocks if $\rho >1/2$,
\item contains no $11$ blocks if $\rho <1/2$,
\item contains neither $00$ nor $11$ blocks if $\rho=1/2$.
\end{itemize}

\section{Rule 232}
Rule 232, also called the ``majority rule,'' has the following rule table:
\begin{eqnarray*}
& & 000 \rightarrow 0,\, 001 \rightarrow 0,\, 010 \rightarrow 0,\,
011 \rightarrow 1, \\ & & 100 \rightarrow 0,\, 101 \rightarrow 1,\,
110 \rightarrow 1,\, 111 \rightarrow 1,
\end{eqnarray*}
which could be also written as
\begin{equation}
s_i^{t+1} \rightarrow
\mbox{majority}(s_{i-1}^t,s_i^t, s_{i+1}^t).
\end{equation}
Let us assume that the initial
configuration includes no $11$ blocks, but at least one $00$ block.
 It is easy to check that the only preimages of $11$ under ${\rm \bf f}_{232}$ are $0110$, $0111$, $1011$, $1101$,
$1110$, and $1111$, and all of them contain at least one subblock $11$. This
means that if the initial configuration contains no $11$ block, then all
subsequent configurations contain no block $11$ either. Consequently,
all entries in the rule table which contain $11$ (i.e., $110$, $111$,
and $011$) do not matter, and we can change them without affecting
the dynamics. Assuming that they are mapped to zero we obtain
a ``simplified'' rule
\begin{eqnarray*}
& & 000 \rightarrow 0,\, 001 \rightarrow 0,\, 010 \rightarrow 0,\,
011 \rightarrow 0, \\ & & 100 \rightarrow 0,\, 101 \rightarrow 1,\,
110 \rightarrow 0,\, 111 \rightarrow 0,
\end{eqnarray*}
which has the code number 32. The following property of rule 32 can
be easily proved by induction:
\begin{proposition}
\label{proprul32}
The block $b \in {\cal B}_{2n+1}$ is the $n$-step preimage of $1$ if, and only
if, $b=101\ldots01$, i.e., it is an alternating sequence of $1$'s and $0$'s
starting with $1$ and ending with $1$.
\end{proposition}

Now, if $L$ is odd, the $[(L-1)/2]$-step preimage of $1$ has to
have the length $L$, so it has to be the entire initial configuration.
 If the entire initial configuration does not have the form required by
 Proposition \ref{proprul32}, it cannot be the $[(L-1)/2]$-step preimage of $1$.
Therefore, after $[(L-1)/2]$ iterations of rule 232 the system converges
to a state of all zeros. For even $L$ this happens after $(L-2)/2$
 iterations. Similarly, if the initial
configuration includes no $00$ blocks, but at least one $11$ block,
the system converges
to a state of all ones. If the initial configuration contains neither
$00$ nor $11$, it stays in this state forever.

Using Propositions 2 and 3, our final result follows immediately:
\begin{proposition}
Let s be a configuration of length $L$ and density $\rho$,
 and let $n=\lfloor (L-2)/2 \rfloor$, $m=\lfloor (L-1)/2 \rfloor$. Then $F_{232}^m(F_{184}^{n}(s))$
consists of only 0's  if $\rho<1/2$ and of only 1's if $\rho>1/2$. If
$\rho=1/2$, $F_{232}^m(F_{184}^{n}(s))$ is an alternating sequence of
0 and 1, i.e., $\ldots01010101\ldots$.
\end{proposition}
As we showed, first $n$ iterations of rule 184 eliminate
all  blocks 11 if $\rho<1/2$ (or 00 if $\rho>1/2$), and the subsequent
$m$ iterations of rule 232 produce homogeneous configuration of
of all 0's (or all 1's). Configurations with $\rho=1/2$ are also treated
properly, i.e., their density remains conserved and the converge
to $\ldots01010101\ldots$. Examples are shown in Figure 1.
%%%%%%%%%%%%%%%%%%%%%%%%%%%%%%%%%%%%%%%%%%%%%%%%%%%%%%%%
% in double column mode, \epsfxsize=2.8in works best
%%%%%%%%%%%%%%%%%%%%%%%%%%%%%%%%%%%%%%%%%%%%%%%%%%%%%%%%
\begin{figure}
\begin{center}
\mbox{    \epsfxsize=3.3in \epsfbox{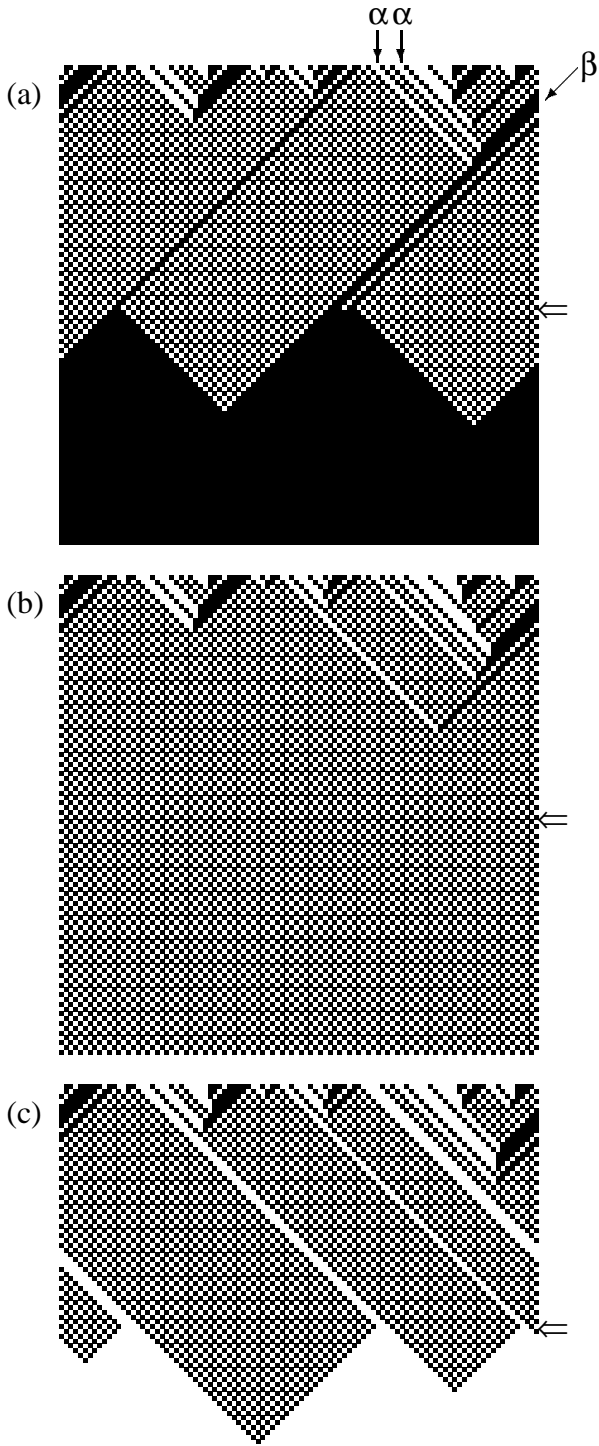}}
\end{center}
\caption{Spatiotemporal diagrams for the (184,232) pair for
lattice size $L=100$ and the initial density a) 0.52 b) 0.5 c) 0.48.
An example of two $\alpha$ defects colliding with the $\beta$ defect is
shown in a). Both rules were iterated 50 time steps (open arrows
indicate where the iteration of rule 232 starts).}
\end{figure}

\section{Remarks}
In conclusion, we have demonstrated that the density classification
task can be performed perfectly in  $L$ time steps with two cellular
automata rules, rule 184 used in the first $L/2$
steps and rule $232$ used in the remaining time steps.
The advantage of this ``assembly line'' processing over a single
rule is evident, as the single rule can never be 100\% successful in
density classification.

The existence of this perfect solution does not mean, of course, that
the evolutionary process could (or could not) produce such a pair of
rules.
Therefore, it would be interesting to design a genetic algorithm
experiment in which pairs of CA rules are evolved, and find out
how easy (or difficult) it is to produce pairs of ``cooperating'' rules
performing better than single rules evolved in earlier experiments.
Since the exact solution exists, we may speculate that
the average performance of a pair obtained in such an experiment should
be significantly better.

Although the solution proposed here performs the task in $L$ time steps,
it is straightforward to construct a faster algorithm, providing that we
allow rules of larger radius. If $f$ is a radius-1 rule, than $f^n$, the
rule iterated $n$ times, is itself a CA rule of radius $n$. Therefore,
the pair $(g,h)$, where $g=f_{184}^n$ and $h=f_{232}^n$, performs the
classification task in $L/n$ time steps, assuming that we iterate both $g$
and $h$ for $L/2n$ time steps.

Another interesting question is the possibility of constructing a general
algorithm to discriminate configurations according to an arbitrary
critical density $\rho_c$. One promising approach to this problem involves
generalized traffic rules, for example rules with higher ``speed limits''
\cite{fukui95},
where the occupied site can move to the right by up to $m$ units if the
sites in front of it are empty. Rules of this type exhibit a phase
transition at $\rho_c=1/(m+1)$ similar to the phase transition in rule 184.
Any configuration with $\rho=\rho_c$
converges to the periodic state of isolated 1's separated by blocks of $m$
zeros. Blocks of zeros longer than $m$ are $\alpha$-type defects,
propagating to the right, while blocks of 1's longer than 1 are defects
of
$\beta$-type, propagating to the left. As in rule 184, $\beta$
defects are eliminated when $\rho < \rho_c$, and $\alpha$ defects
disappear when $\rho > \rho_c$.
One can also construct an analog of
rule 232 which grows $\alpha$ defects if $\beta$ defects are not
present and conversely, and such a pair of radius-$m$ rules
can perform the $\rho_c=1/(m+1)$ classification task
for any integer $m>0$ (details of this construction will be presented
elsewhere). It is not clear, however, how to generalize
this method for arbitrary $\rho_c$.

\section{Acknowledgements}
 The author is grateful to Prof. Nino Boccara
for  useful discussions and reading of the manuscript.

\end{document}